\documentclass[conference]{IEEEtran}

\usepackage{times}
\usepackage{epsfig}
\usepackage{graphicx}
\usepackage{amsmath}
\usepackage{amssymb}
\usepackage{rotating}
\usepackage{multirow}
\usepackage{booktabs}
\usepackage{ctable}
\usepackage{colortbl}
\usepackage{etex}
\usepackage{url}
\usepackage{graphicx}

\begin{document}
	
\title{The Horcrux Protocol: A Method for Decentralized Biometric-based Self-sovereign Identity}
\author{%
	% author names are typeset in 11pt, which is the default size in the author block
	{Asem Othman and John Callahan}%
	% add some space between author names and affils
	\vspace{0.1mm}\\
	\fontsize{9}{9}\selectfont\ttfamily\upshape
	Veridium IP Ltd\\
	aothman@veridiumid.com, jcallahan@veridiumid.com%
}

\maketitle
\begin{abstract}
	
Most user authentication methods and identity proving systems rely on a centralized database. Such information storage presents a single point of compromise from a security perspective. If this system is compromised it poses a direct threat to users’ digital identities. This paper proposes a decentralized authentication method, called the Horcrux\footnote{The term ``horcrux'' comes from the Harry Potter book series in which the antagonist (Lord Voldemort) places copies of his soul into physical objects. Each object is scattered and/or hidden to disparate places around the world.  He cannot be killed until all horcruxes are found and destroyed.} protocol, in which there is no such single point of compromise. The protocol relies on decentralized identifiers (DIDs) under development by the W3C Verifiable Claims Community Group and the concept of self-sovereign identity. To accomplish this, we propose specification and implementation of a decentralized biometric credential storage option via blockchains using DIDs and DID documents within the IEEE 2410-2017 Biometric Open Protocol Standard (BOPS).
	 
\end{abstract}

\section{Introduction}

Digital transformation, mobility and the proliferation of applications and networks have made traditional forms of information protection increasingly difficult to manage and enforce. Information is everywhere, access is widely distributed, but most security programs are still largely based on archaic, static models that just don’t work anymore and it is getting worse. 

The latest evidence of this is recent breach disclosed by Equifax \cite{hume2004identity} that has exposed identity information for over 140 million individuals. Enterprises continue to take on enormous risk by aggregating unnecessary personal data while customers can’t manage the massive number of IDs, passwords and data required to interact with every on-line connection.

We believe that the common denominator across most aspects of information protection is identity. An identity is inextricably linked to a person, device, application, system or network and it is the most dependable ‘perimeter’ we can rely upon to determine how to make information available properly and securely. Identity management will soon have to make the leap from our age-old approaches of multiple user IDs and passwords to a new, secure, privacy-centric means of identity authentication.

An identity ecosystem leverages personas that can both protect privacy (and reduced liability for the enterprise), provide distributed access to authorized services and provide the user a full-control over their identity accessing.  User authentication presents one of the basic security requirements in this identity ecosystem. Generally speaking, authentication can be described as a process in which a user offers some form of proof that he is the same user who registered the account. A proof of identity can be any piece of information that an authentication server accepts: something users have in their possession, something they know or something they are (e.g., a biometric).

\subsection{Traditional Authentication models}

In current practice, only one centralized database is in charge of storing the data used for authentication. When the user offers the requested proof of identity, the authentication server evaluates this proof and grants access to the user. For example, when a user tries to access his account on a typical web application he is prompted to enter a password. Traditionally, the web application holds the information about the user’s account and his password. When the user submits his password during log-in process, the application compares the stored password to the submitted password. If they match, the user is granted access to the application. In other words, all the information needed to authenticate the user is held on a single system. 
Even if the authentication system is biometric-based system, most of the deployed systems is still use the same centralized model.

Biometric-based authentication systems \cite{jain2004introductiontobiometrics} operate in two main stages: enrollment and recognition. The enrollment stage generates a digital representation of an individual's biometric trait and then stores this representation called biometric template in a centralized system database. During the recognition stage, which can be operate in two modes: verification and identification, the system require that the acquired probe biometric template to be matched against a single template (in the verification mode) or all template (in the identification mode) stored in the centralized database.

This makes such systems the single point of compromise for securing digital identities. In other words, in case an attacker gains access to the web application or the biometric centralized database, he can extract enough information to compromise the user’s digital identity \cite{ jain2008biometric}. Moreover, since many users tend to use the same password or biometric trait in different applications, revealing their identity on one compromised database can lead to unlawful access into other accounts and services.

In some current implementations, the authentication server can be completely separated from the server running web applications or biometric authentication database . For example, single sign-on (SSO) schemes \cite{radha2012survey} are based on this concept. SSO schemes rely on a third-party identity provider (IdP) to broker authentication using protocols such as SAML \cite{hughes2005security} and OpenID Connect \cite{sakimura2011openid}.  Since their introduction in 2002 and 2010 respectively, only 5\% of sites use any of over 50 disparate IdP \cite{vapen2016look} SSO services (e.g., ``login with Facebook'', ``login with Google'', etc.).  Loopholes in these centralized IdP-based SSO systems are the main reasons for the many hacks of personal information \cite{hume2004identity} and even loss of biometric data \cite{zetter2015opm}. Surveys of users show an overwhelming dissatisfaction with single-sign-on (SSO), a feeling of ``lack of control'' over their data \cite{innovalor2016,mertens2015digital,rose2012value,satchell2011identity} and a desire to control it themselves.  Upcoming legislation, such as the General Data Protection Regulations (GDPR) \cite{koops2014privacy} and Payment Services Directive II (PSD2) \cite{cortet2016psd2}, are pressuring institutions, both private and public, to place citizen or customer data into the end user's control.

\subsection{Traditional Identity Proving Methods}

\begin{figure}[htbp] \begin{center}
		\includegraphics[width=0.3\textwidth]{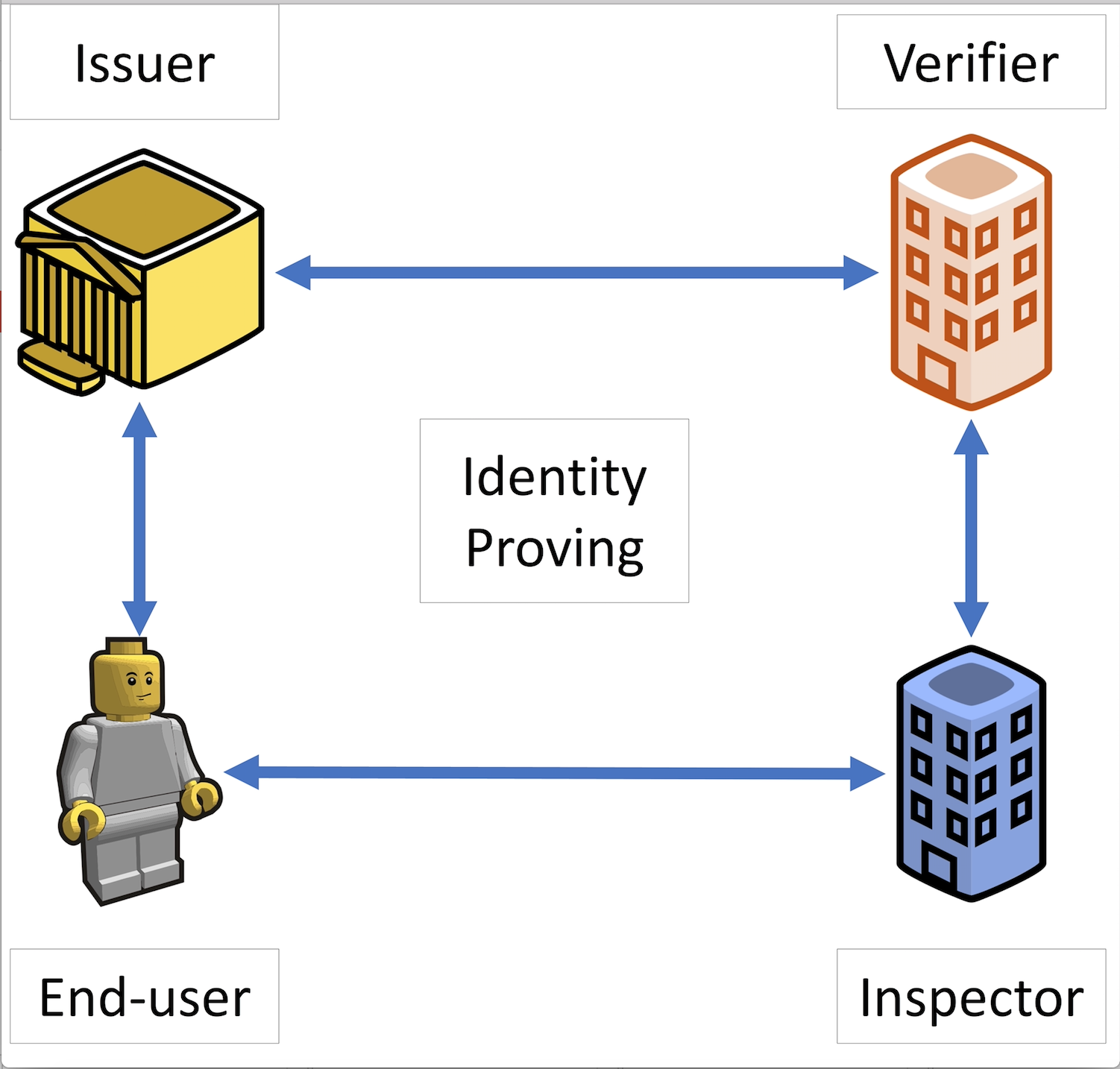}
		\caption{\label{fig:OldEcoSystem} Traditional Identity Proving Ecosystem.}
\end{center} \end{figure}

Current identity proving methods  (see Figure \ref{fig:OldEcoSystem}) rely on specific parties: an {\em issuer}, {\em end-user}, {\em verifier}, and {\em inspector}.

Issuers such as governments associate identity credentials to end-users. Then, the issuer shares personal information and credentials of the end-user with a verifier. If the end-user applies for a bank account, credit card, or car loan, the inspector contacts a verifier to prove the claimed identity by the end-user. Therefore, especially if this process is online, the inspector presents a multiple-choice quiz about past addresses or who financed the user's last car. That’s an identity verification service that verifier provides to lenders and others, i.e., inspectors. Based on the answers or prove of holding the credentials, the inspector will verify the claimed identity by the end-user and grantee the required service. This ecosystem has the same security flaw as the traditional authentication systems, end-user personal data (e.g., SSN, addresses, birthdate, etc.) are stored in a centralized database of the verifier.

\subsection{Our Contribution}

The aforementioned security flaws encapsulate perfectly why a new identity ecosystem is so important: identity is the new attack surface \cite{los2016}. In traditional authentication and identity models, users are forced to relinquish personal information such as credit histories, credentials such as birth certificate, or biometric data such fingerprint template to a third party, with a centralized database. 

{\em Self-sovereign identity} is a new decentralized ecosystem for private and secure identity management that is being implemented by several projects \cite{reed2017beyond,ali2016blockstack,lundkvist2016uport} as the replacement of the traditional identity proving systems. Self-sovereign identity puts end-users — not the organizations that traditionally centralize identity — in charge of decisions about their own privacy and disclosure of their personal information and credentials. Self-sovereign identity utilizes distributed ledgers, i.e., blockchain technology, to establish a web-of-trust \cite{wot}. These blockchains are a form of databases that is provided cooperatively by a set of organizations, instead of by a central database with a central organization. A single blockchain is copied redundantly in many places, and it accrues transactions orchestrated by many machines. In other words, the new identity model is a reliable, public identity proving system under no single entity’s control, robust to system failure and hacking.

In this paper, we discuss the specification and implementation of our Horcrux protocol that combines the decentralized self-sovereign identity ecosystem with 2410-2017 IEEE Biometric Open Protocol Standard (BOPS)\cite{BOPS-2017}.  The BOPS protocol is extensible to a combination of on-device (FIDO UAF \cite{FIDO-UAF-2014} compatible), server-side or a multi-distribution model that utilizes a secret scheme.  Indeed, the standard allows for off-device biometric credentials under user control.  The device’s local TPM is only one option (though dominant at the moment) for persisting biometric credentials and associated key(s). 

The Horcrux protocol allows the end-users of self-sovereign identity to have the control of accessing their identities by giving the consent to this verification process via a biometric authentication process. Moreover, We propose the use of the existing BOPS due to its multi-distribution scheme of storing biometric data. BOPS utilizes a secret scheme to divide the templates into $n \leq 2$ shares as specified in IEEE 2410-2017. Therefore, biometric data used for authentication will be distributed by BOPS and securely stored in decentralized storages and securely referenced to them by blockchains technology. The multiple shares (and potentially redundant shares) could be spread across alternate off-chain storage (like IPFS, Dropbox, Google drive, etc.) as designed in the self-sovereign ecosystem. 

This marriage of these two identity models (DIDs and BOPS) is the Horcrux protocol which guarantees the following principles:

\begin{itemize}
	\item {\em Existence}: users must have an independent existence that can not only exist wholly in the digital form, and by using biometric-based protocol for enrolling and authentication, this guarantees that the digital identity has been created and will always be verified by an existence end-user.
	\item {\em Control}: users must control the storage and access to their identities. Under the Self-sovereign identity ecosystem, users always able to refer to, update, or even hide their personal information and credentials. Our Horcrux protocol will assure that the access is always secure by their biometric which also is securely stored via the decentralized ecosystem, along with their personal information.
	\item {\em Portability and interoperability}: BOPS and self-sovereign identity have been designed around these principle.
	\item {\em Protection}: the security of Horcrux protocol is trusted because it is based on strong cryptography and governed by self-sovereign identity via a blockchain technology and BOPS.
\end{itemize}

The rest of the paper is organized as follows. Sections \ref{sec:BOPS} and \ref{sec:sovereign} present IEEE Biometric Open Protocol Standard (BOPS) and Self-sovereign identity ecosystem, respectively. Section \ref{sec:horcrux} discuss our Horcrux protocol and its implementation. Finally, Section \ref{sec:con} summarizes the paper.

\section{BOPS}\label{sec:BOPS}

Biometric authentication demands high assurance levels such as those required by national and international standards \cite{nist800633}.  The IEEE 2410-2017 Biometrics Open Protocol Standard (BOPS) \cite{BOPS-2017} defines the following elements to achieve required levels of assurance:

\begin{itemize}
	\item {\em Collection}: BOPS defines application programming interfaces (API) such that biometric templates (fingerprints, facial, voice, etc.) are collected via a hardware security module (HSM), trusted execution environment (TEE) or trusted platform module (TPM) when possible.  Such facilities ensure non-accessible and/or encrypted memory to prevent exfiltration of biometric data.
	\item {\em Storage}: BOPS defines secure formats and envelopes such that biometric data persisted via encryption using a hardware security module (HSM), trusted execution environment (TEE) or trusted platform module (TPM) when possible.  Such facilities ensure non-accessible and/or encrypted memory to prevent exfiltration of biometric data.  BOPS also accommodates methods for cryptographic sharding \cite{ross2011visual} such that a share is kept locally on the device and a second share can be kept locally or sent to the remote platform.  Loss of either share does not compromise the complement share nor the biometric template.
	\item {\em Transmission}: BOPS defines a Representational state transfer (REST) interface protocol such that no biometric is transmitted unless it is encrypted in within an envelope using the server's public key (per enrollment) over a two-way TLS channel.
	\item {\em Processing}: BOPS requires matching of biometric templates in volatile memory or using the local HSM, but never persisted to any form of non-transient storage such as files, databases, or other long-term storage media.
\end{itemize}

BOPS defines two phases of operation: enrollment and authentication.  During enrollment, the remote server generates a public-private key pair (RKP) in which the public key is sent to the mobile device.  Then, a biometric template (called the initial biometric vector or "IBV") is collected  and paired with a device-generated public-private key pair (LKP) using the local HSM when available.  The LKP private key is reserved locally and the LKP public key along with the biometric share(s) are encrypted with the RKP public key for transmission to the server over a two-way TLS connection.  The client certificate for the TLS connection is installed a priori via application installation on the mobile device.  

Biometric authentication requires collection of a candidate biometric vector (CBV) for comparison to the IBV.  BOPS defines three configuration modes for authentication:

\begin{itemize}
	\item {\em Local}: The collected CBV is compared on the device to the reconstructed IBV shares.  The match result can be a threshold value or a boolean that is encrypted in an envelope using the RKP public key and transmitted to the server.  This mode is FIDO UAF \cite{FIDO-UAF-2014} compliant when used with a certified local FIDO UAF authenticator.
	\item {\em Remote}: The collected CBV is encrypted in an envelope with the RKP public key and transmitted to the server for comparison on the remote server.
	\item {\em Local Match}: The server is requested to encrypt (using its RKP private key) any IBV shares it holds and return them to the local device.  The CBV is collected, IBV share(s) from local and remote combined and matched on the local device.  The CBV and combined IBV are subsequently wiped from volatile memory.
	\item {\em Remote Match}: The collected CBV and any local IBV share(s) are encrypted in an envelope with the RKP public key and transmitted to the server.  On the server, the incoming IBV share(s) from the local device are combined with server-based share(s) and compared to the incoming CBV.
\end{itemize}

The BOPS protocol also uses one-time password and server-based challenges in envelopes to prevent man-in-the-middle (MITM) and replay attacks that might threaten the security of biometric data and other credentials in transit.  A recent comparison \cite{oxfordmc2017} shows that FIDO UAF and BOPS offer rough comparable protection against such threat vectors.  In Local configuration mode, BOPS and FIDO UAF are comparable, but BOPS offers additional modes for remote (and sharded) storage and matching.  Remote storage and match of biometric data may not be appropriate in some jurisdictions and regulatory regimes, but it depends on each institution's policies, cyber security standards, risk compliance levels and assurance needs.

\section{Self-sovereign identity ecosystem}\label{sec:sovereign}
Self-sovereign identity is a new identity ecosystem where individuals (or even organization) to whom the identity pertains, control and manage their identities. In this sense the individual is their own identity provider—no external party can claim to “provide” the identity for them because it is intrinsically theirs. In other words, self-sovereign identity is as a digital record or container of identity transactions that end-users control. The end-user can add more data to it, or ask others to do so, reveal some the data or all of it some of the time or all the time.

Moreover, end-users can record their consent to share data with others, and easily facilitate that sharing. It is persistent and not reliant on any single third party. Claims made about an end-user in identity transactions can be self-asserted or asserted by a 3rd party whose authenticity can be independently verified by a relying party. The infrastructure of self-sovereign identity has to reside in an environment of diffuse trust which is not controlled by any single organization or even a small group of organizations. The cryptographically secure blockchain is the breakthrough technology that makes this possible. It enables multiple entities such as organizations and governments to cooperate mutually via distributed consensus to form decentralized blockchains, where data is replicated in multiple locations to be resistant to faults and tampering. While distributed ledger technology has been around for some time, new blockchain applications, such as Bitcoin, have resulted in realizations of its potential, particularly with respect to decentralization and security.

\begin{figure}[htbp] \begin{center}
		\includegraphics[width=0.3\textwidth]{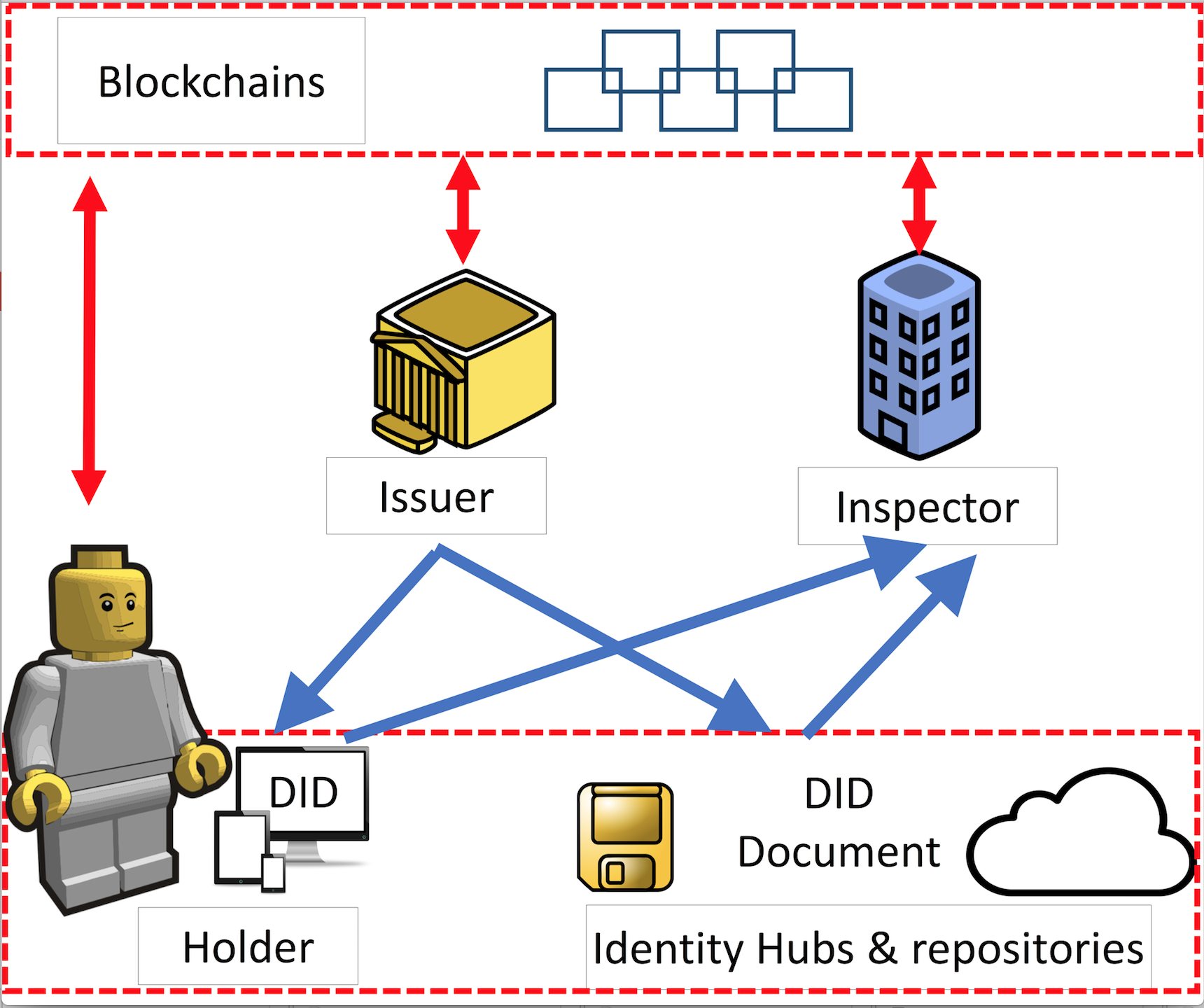}
		\caption{\label{fig:newEcoSystem} Self-sovereign Identity Ecosystem architecture.}
\end{center} \end{figure}

Figure \ref{fig:newEcoSystem} provides an overview of the self-sovereign identity architecture. The followings are the brief descriptions of the architecture entities. Note that in this architecture, the information is no longer centralized and connections are individually permissioned.

\begin{itemize}
	\item {\em DID:} Decentralized Identifiers (DIDs) are a new type of identifier intended for a self-sovereign identity system, i.e., entirely under the control of an entity and not dependent on a centralized registry or certificate authority. DIDs are opaque, unique sequences of bits, that get generated when a user accepts a claim from an issuer along with a corresponding DID Document. DIDs have a foundation in (Universal Resource Identifiers) URIs\cite{mealling2002report,didspec1}; therefore, they achieve global uniqueness without the need for a central registration authority.

	\item {\em DID document:} A DID resolves to an corresponding DID Document --- a simple document that contain all the metadata needed to interact with the DID. Specifically, a DID Document typically contains at least three things along with personal information or credentials. The first is a set of mechanisms that may be used to authenticate as a particular DID (e.g., public keys, biometric templates, or even encrypted share of biometric data). The second is a set of authorization information that outlines which entities may modify the DID Document. The third is a set of service endpoints, which may be used to initiate trusted interactions with an entity\cite{didspec1}.
	
	\item {\em Blockchains:} In this architectural construct, the blockchain acts as an index of identifiers and audit trail of permissioned exchanges between the issuer of claims, the holder of claims, and the inspector of claims.

	\item {\em Identity hubs and repositories:} These hubs are secure personal data repositories that curate and coordinate the storage of signed/encrypted DID documents, and relay messages to identity-linked devices. Examples of identity hubs include Dropbox, Google drive, and Storj.

	\item {\em Issuer:} An entity that creates DID and DID documents, associates it with a particular subject and transmits it to a holder. Examples of issuers include corporations, governments, and individuals.

	\item {\em Inspector/Verifier:} Inspectors request claims in the form of DIDs from subjects and organizations in order to give them access to protected resources. The inspector verifies that the credentials provided via DID and in the DID document are fit-for-purpose, also checks the validity of the DID in the blockchain. Examples of inspectors include employers, security personnel, and websites.

	\item {\em Holder:} Holders receive DIDs from issuers, store DID Documents via identity hubs, and provide DID Documents to inspectors. The entity which controls a particular DID can be the subject of the DID document, but not necessarily. An inspector can also resolve DIDs into their corresponding DID documents and discovery DIDs across a decentralized system. Examples of holders are users --- students, employees, and customers. Other examples of holders that have the permissions to handle subject’s claims include web services or mobile apps installed on the subject’s personal devices.
\end{itemize}

\section{The Horcrux Protocol}\label{sec:horcrux}

\begin{figure*}[t]
	\centering
	\includegraphics[width=0.7\linewidth]{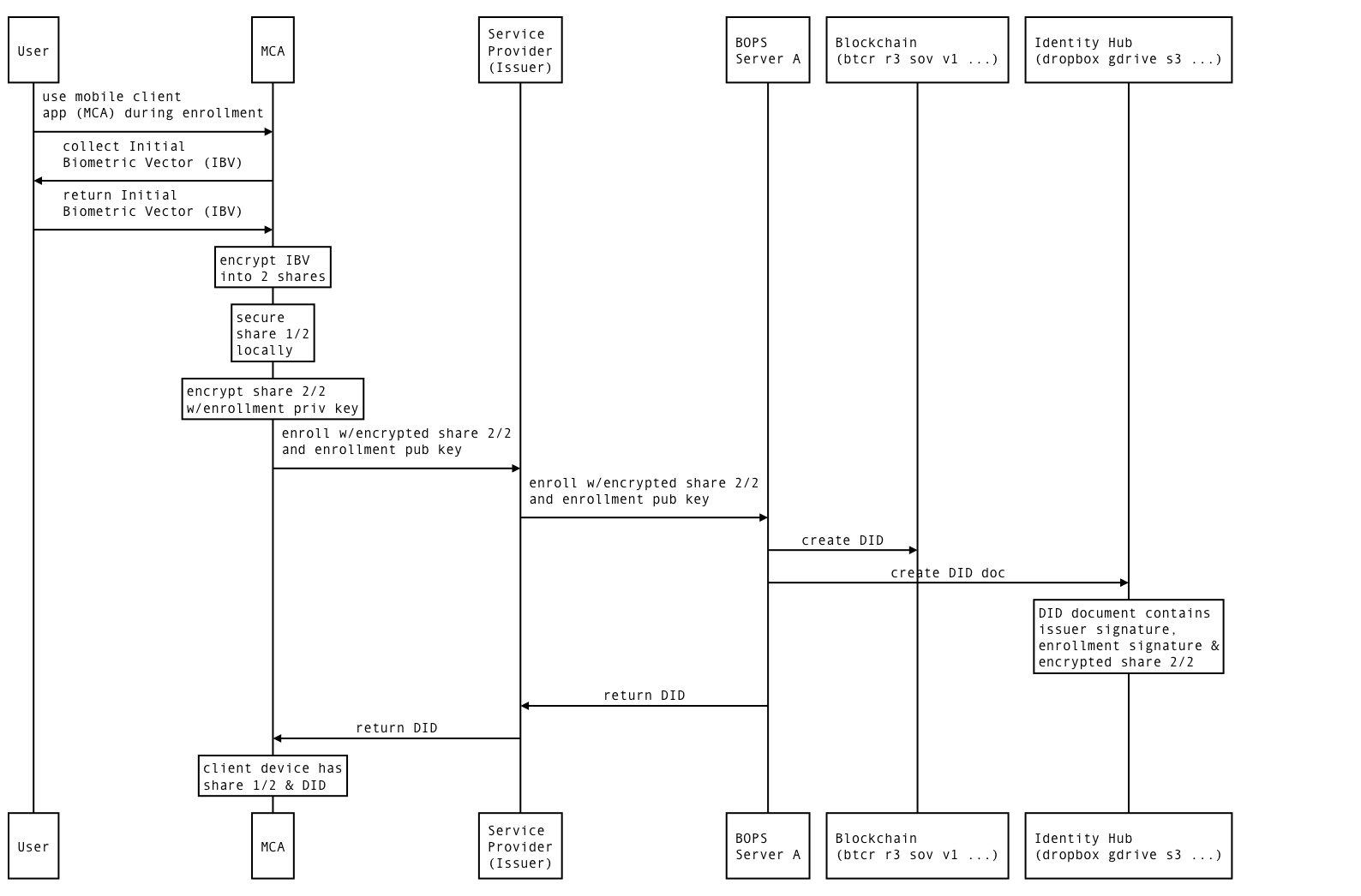}
	\caption{Enrollment sequence}
	\label{fig:enrollhorcrux}
\end{figure*}

The IEEE 2410-2017 standard allows for interoperablility at several layers including the persistence cluster (\cite{BOPS-2017} section 7.3.3) provided it satisfies security requirements for storage of encrypted biometric shares.  We propose any BOPS server can act as a {\em holder} of biometric shares via blockchain using methods outlined in the W3C Decentralized Identity (DID) specification\cite{didspec1}. A BOPS server can enroll a user by storing biometric share(s) as DID Documents using off-chain storage providers owned by the user.  The corresponding DID acts as the identity assertion associated with the enrolled biometric.  Figure~\ref{fig:enrollhorcrux} depicts a standard BOPS enrollment flow (adapted from \cite{BOPS-2017} section 7.2).  The user (via a browser user-agent) is prompted to enroll their biometrics with a service provider acting as an {\em issuer}.  The initial biometric vector (IBV) is encrypted (via visual cryptopraphy) into two shares.  One share is reserved on the local mobile device while the second is transmitted to the BOPS server.  Instead of an RDBMS or persistence cluster (e.g., SOLR) backend, the BOPS server relies on a blockchain store in this case using a decentralized identitifer (DID)\cite{didspec1} for persistence. DIDs provide a blockchain-agnostic method for resolving DID Documents much like URIs \cite{mealling2002report} uniquely characterize web resources via URNs and URLs, but for disparate blockchain ecosystems.  The W3C Verifiable Claims Community Working Group has defined DID method specifications \cite{didspec1} for implementors of CRUD operations specific to a particular blockchain.  The BOPS server acts as a resolver given a DID to fetch the corresponding DID Document if possible.  The DID and corresponding DID Document are cryptographically associated with each other via blockchain transactons such that any tampering with the DID Document for a given DID would be evident.  After persisting the DID document and registering the associated DID on a blockchain, the user is notified of success (or failure) of their enrollment.  It should be noted that no biometric shares are stored on any blockchains, only in DID Documents that are persisted ``off-chain'' via identity hubs or personal storage providers.

\begin{figure*}
	\centering
	\includegraphics[width=0.7\linewidth]{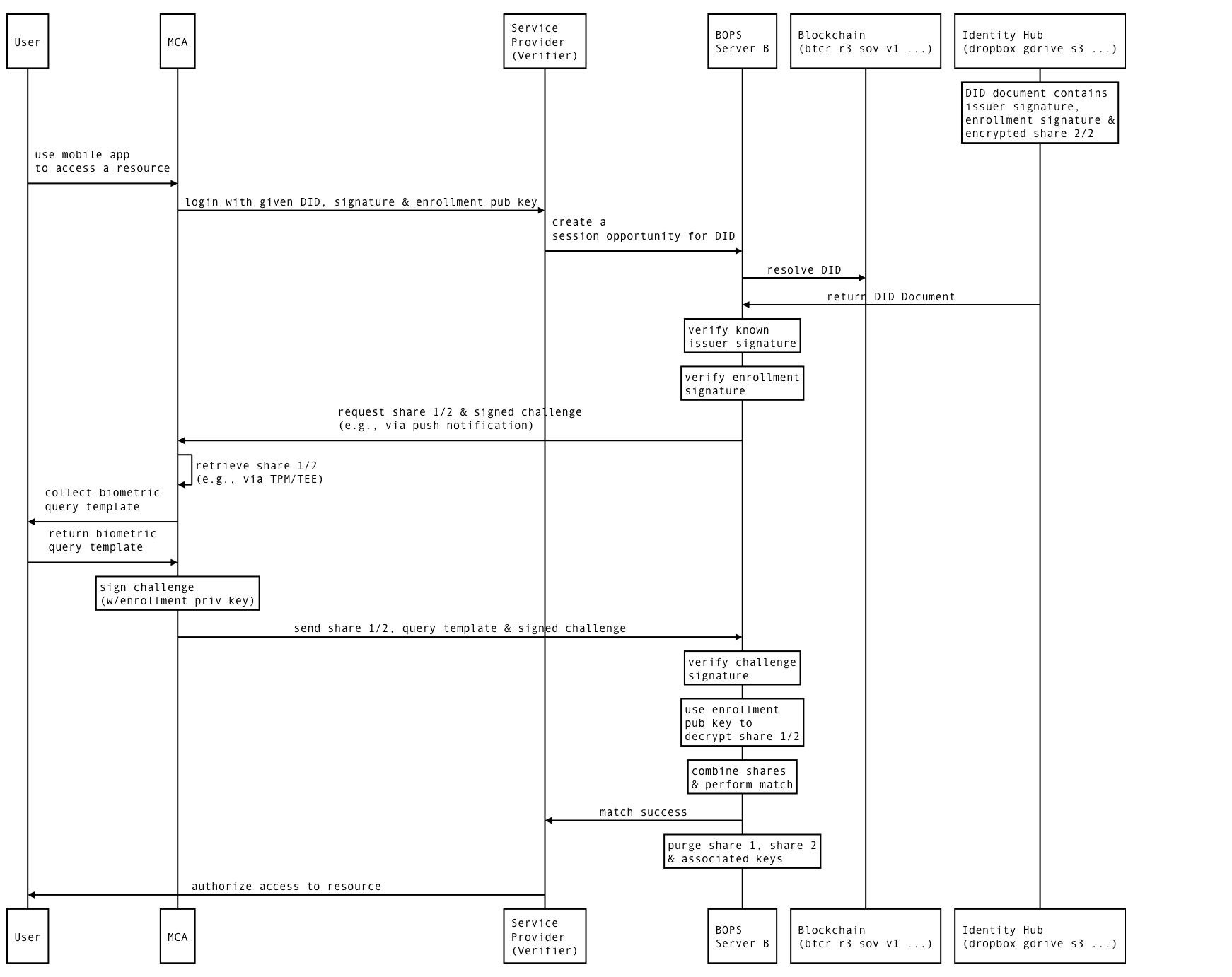}
	\caption{Remote authentication sequence}
	\label{fig:authhorcrux}
\end{figure*}

The encrypted biometric share is still within an encrypted envelope as per \cite{BOPS-2017} but the share is persisted on a corresponding blockchain with an associated DID.  The DID can be used as a claim with another BOPS server acting as a {\em verifier}.  Again, this is possible because any tampering with the DID Document associated with a given DID will be detectable due to their relationship via a recorded blockchain transaction\cite{didspec1}.  Figure~\ref{fig:authhorcrux} shows an example of a different BOPS server being used by a verifier.  In this example, the user tries to access a resource on a web site (e.g., the service provider) using a mobile client application (MCA) with a DID created by an issuer (\ref{fig:enrollhorcrux}) and a public key created at enrollment.  The service provider relies on a BOPS server to resolve the DID and fetch the corresponding DID Document via a blockchain from the storage provider.  If the DID document is a valid claim, the BOPS server checks if the issuer of the claim is known (via its public key in the DID document) and that the enrollment public key matches for this user as well.  If valid, the user (via their MCA) is requested for their candidate biometric vector (CBV) and complement share of the IBV as per \cite{BOPS-2017}.  Upon receiving the complementary share and CBV from the client (as described in \ref{sec:BOPS} - Remote configuration mode), the enrollment public key is used to decrypt the client's share, combine the IBV shares and match them to the CBV.  If successful, the user is authenticated.

In the case of remote authentication, the service provider, acting as a verifier, uses a different BOPS server instance to authenticate the user even though this user has never registered at this service provider.  Furthermore, the user and service provider are the only parties needed at authentication time unlike SAML or OAuth that rely on 3rd party identity providers (IdPs) to broker identity claims in traditional single-sign-on (SSO) systems.  The Horcrux protocol supports {\em self-sovereign identity} \cite{baars2016towards} by using blockchain technology to secure credentials issued by valid authorities (i.e., {\em issuers}) for later use directly by the user who owns the credentials.  The user may store such credentials via several personal cloud storage providers such as Dropbox, Google drive, Amazon S3, etc. but delegate management (via OAuth tokens) to a {\em holder} such as the BOPS server.  The holder can access issued claims like the ecnryoted biometric shares on behalf of the user during authentication, but require biometric authentication as specified in the {\tt authenticationCredentials} section of the claim \cite{didspec1}.

\begin{figure*}
	\centering
	\includegraphics[width=0.7\linewidth]{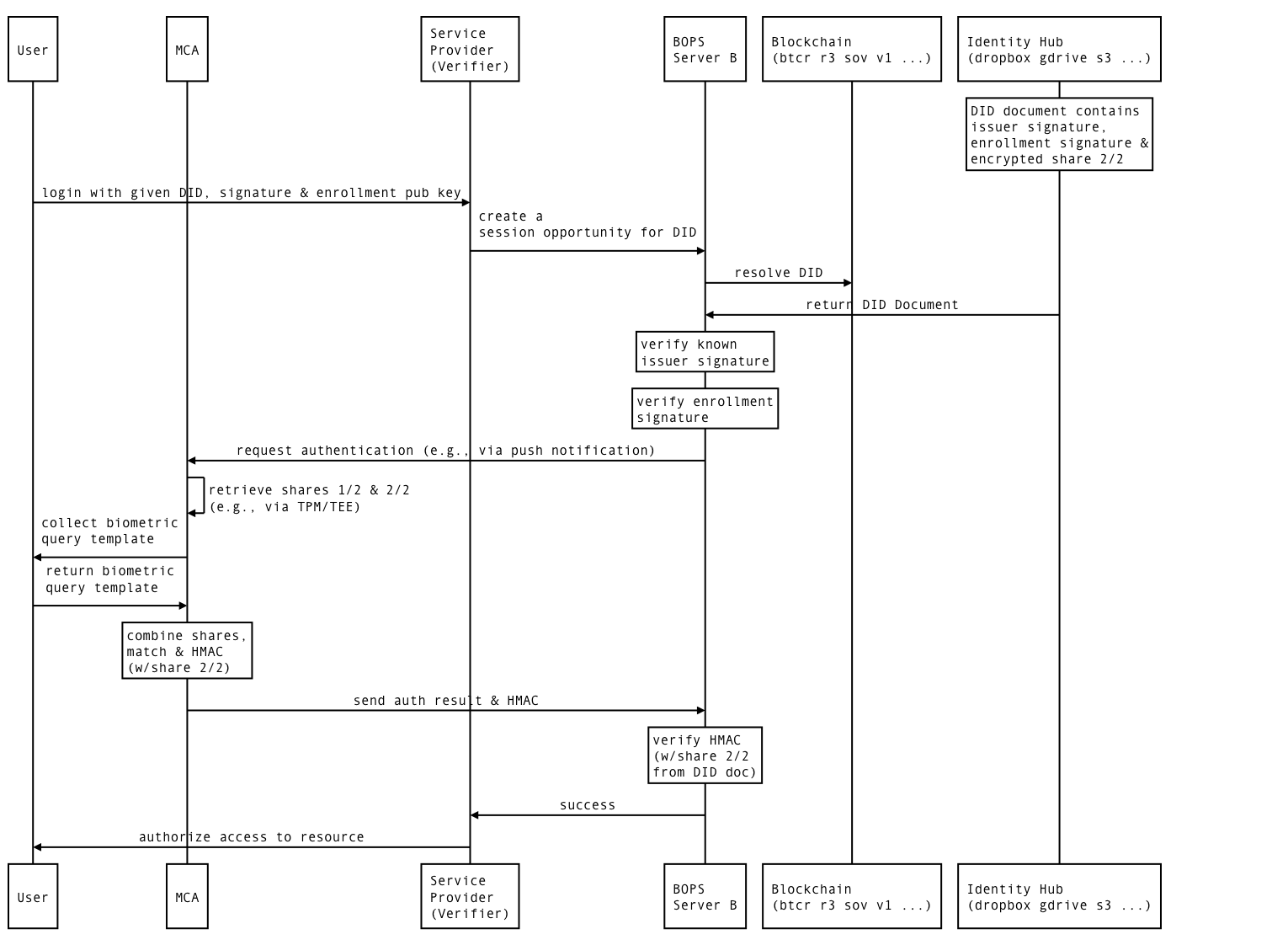}
	\caption{Local authentication sequence}
	\label{fig:localhorcrux}
\end{figure*}

The local configuration mode of BOPS is also available such that a combination of biometric shares occurs on the mobile device.  Figure~\ref{fig:localhorcrux} shows this variation in which the second biometric share is retreived via DID referencing from the corresponding DID document but transmitted to the client by a service provider and its BOPS server.  The biometric share is opaque to the service provider and BOPS server in this case, but the server knows that the corresponding share on the mobile device is used for matching due to the HMAC of the encrypted second share.  The enrolled share is never sent to the device, but both shares are kept locally as per BOPS local configuration mode.  The mobile device must hold the private key associated with the enrolled share for the DID because it computes an HMAC using the share and sends it to the server.  The server can compare the HMAC key with the opaque encrypted share from the DID document.  It is possible, however, that the user could resolved a given DID, retrieve the corresponding DID document, extract the opaque encrypted share and construct the HMAC thus spoofing possession of that share and falsifying the biometric match.  We are in the process of investigating methods for securing DIDs on a mobile device and/or using server-based key mechanism to prevent this attack vector.

The IEEE 2410-2017 standard allows for more than two encrypted shares.  Algorithms such as visual cryptography \cite{ross2011visual} and Shamir secret sharing \cite{naor1994visual} allow for larger number of shares that.  Using DIDs and associated DID documents for more biometric shares across different blockchains and replicating copies of shares could further protect users from compromise and increase availability.

\section{Summary}\label{sec:con}

The self-sovereign identity model provides authority-based issuance of claims and eliminates the need for 3rd-party identity providers during authentication using blockchain technologies to assure exchange of verifiable credentials. The Horcrux protocol is a method for secure exchange of biometric credentials within an existing standard (IEEE 2410-2017 BOPS \cite{BOPS-2017}) implemented across next-generation blockchain-based self-sovereign identity platforms based on open standards like DIDs and DID Documents \cite{didspec1}.  By using blockchain and off-chain storage as an alternative to the persistent layer in BOPS, we use new blockchain-agnostic standards to enroll via an issuer and authenticate on a verifier that are not part of an real-time trust network.  Instead, they rely on user-controlled biometric credentials that are cryptographically encrypted into multiple shares across the user's device and blockchain-linked personal storage providers.  The protocol is generalized for two or more biometric shares that can be stored across mobile devices and personal storage providers with redundancy for availability and safety. Future plans include a reference implementation and detailed analysis of the protocol for performance and correctness using TLA+ in a manner similar to the protocol analysis of WPA found in \cite{narayana2006automatic}.

%\section{REFERENCES}
\tiny{
\label{sec:ref}
\bibliographystyle{ieee}
\bibliography{New_ref_bib}
}
\end{document}